\documentclass[twocolumn,english,aps,prb,floats,showpacs]{revtex4}
\usepackage[T1]{fontenc}
\usepackage[latin9]{inputenc}
\usepackage{bm}
\usepackage{amsmath}
\usepackage{amssymb}
\usepackage{graphicx}
\usepackage{esint}

\makeatletter
\@ifundefined{textcolor}{}
{%
 \definecolor{BLACK}{gray}{0}
 \definecolor{WHITE}{gray}{1}
 \definecolor{RED}{rgb}{1,0,0}
 \definecolor{GREEN}{rgb}{0,1,0}
 \definecolor{BLUE}{rgb}{0,0,1}
 \definecolor{CYAN}{cmyk}{1,0,0,0}
 \definecolor{MAGENTA}{cmyk}{0,1,0,0}
 \definecolor{YELLOW}{cmyk}{0,0,1,0}
}

\@ifundefined{textcolor}{}{%
 \definecolor{BLACK}{gray}{0}
 \definecolor{WHITE}{gray}{1}
 \definecolor{RED}{rgb}{1,0,0}
 \definecolor{GREEN}{rgb}{0,1,0}
 \definecolor{BLUE}{rgb}{0,0,1}
 \definecolor{CYAN}{cmyk}{1,0,0,0}
 \definecolor{MAGENTA}{cmyk}{0,1,0,0}
 \definecolor{YELLOW}{cmyk}{0,0,1,0}
}

\usepackage{babel}
\usepackage{bm}

\makeatother

\usepackage{babel}
\begin{document}

\title{Writing Skyrmions with a Magnetic Dipole}

\author{Dmitry A. Garanin$^{1}$, Daniel Capic$^{1}$, Senfu Zhang$^{2}$,
Xixiang Zhang$^{2}$, and Eugene M. Chudnovsky$^{1}$}

\affiliation{$^{1}$Physics Department, Herbert H. Lehman College and Graduate
School, The City University of New York, 250 Bedford Park Boulevard
West, Bronx, New York 10468-1589, USA \\
 $^{2}$Physical Science and Engineering Division (PSE), King Abdullah
University of Science and Technology (KAUST), Thuwal 23955-6900, Saudi
Arabia}

\date{\today}
\begin{abstract}
We demonstrate numerically on large spin lattices that one can write
skyrmions in a thin magnetic film with a magnetic dipole of a few
tens of nanometer in size. Nucleation of non-chiral skyrmions as well
as chiral skyrmions formed by the Dzyaloshinskii-Moriya interaction
has been investigated. Analytical model is developed that agrees with
numerical results. It is shown that skyrmions can be written though
a number of scenarios that depend on the experimental technique and
parameters of the system. In one scenario, that branches into subscenarios
of different topology, the magnetic dipole on approaching the film
creates a skyrmion-antiskyrmion pair. As the dipole moves closer to
the film it induces collapse of the antiskyrmion and creation of a
non-zero topological charge due to the remaining skyrmion. In a different
scenario the dipole moving parallel to the film nucleates a skyrmion
at the boundary and then drags it inside the film. Possible implementations
of these methods for writing topologically protected information in
a magnetic film are discussed. 
\end{abstract}

\maketitle

\section{Introduction}

Skyrmions in thin films are currently at the forefront of theoretical
and experimental research in magnetism due to their potential for
topologically protected information storage and logic devices \cite{Nagaosa2013,Zhang2015,Klaui2016,Leonov-NJP2016,Hoffmann-PhysRep2017,Fert-Nature2017}.
Research in this area has focused on skyrmion stability, dynamics
and various symmetry properties. Anisotropy, dipole-dipole interaction
(DDI), magnetic field, and confined geomery can stabilize significantly
large magnetic bubbles with skyrmion topology \cite{IvanovPRB06,IvanovPRB09,Moutafis-PRB2009,Ezawa-PRL2010,Makhfudz-PRL2012},
while stability of small skyrmions requires other than Heisenberg
exchange coupling, strong random anisotropy, or a non-centrosymmetric
system with large Dzyaloshinskii-Moriya interaction (DMI) \cite{AbanovPRB98,Bogdanov-Nature2006,Heinze-Nature2011,Leonov-NatCom2015,Chen-APL2015,Boulle-NatNano2016,Lin-PRB2016,Leonov-NJP2016,EC-DG-NJP2018}.

With an eye on a skyrmionic memory and data processing one of the
most challenging tasks in this field is writing and manipulating skyrmions
in a magnetic film. In a film with perpendicular anisotropy multiskyrmion
topological structures randomly evolve from stripe domains on increasing
the normal component of the magnetic field \cite{GCZ-EPL2017}. For
practical applications one has to be able to generate and manipulate
individual skyrmions. It has been demonstrated that skyrmions can
be created, annihilated and moved by current-induced spin-orbit torques
\cite{Yu-NanoLet2016,Fert-Nature2017,Legrand-Nanolet2017}. Individual
skyrmion bubbles can also be generated by pushing elongated magnetic
domains through a constriction using an in-plane current \cite{Jiang-Sci2015,Hoffmann-PhysRep2017}.
Small skyrmions can be written and deleted in a controlled fashion
with local spin-polarized currents from a scanning tunneling microscope
\cite{Romming-Sci2013}. It has been also shown that light-induced
heat pulses of different duration and energy can write skyrmions in
a magnetic film in a broad range of temperatures and magnetic fields
\cite{Berruto-PRL2018}.

\begin{figure}[ht]
\centering{}\includegraphics[width=9cm]{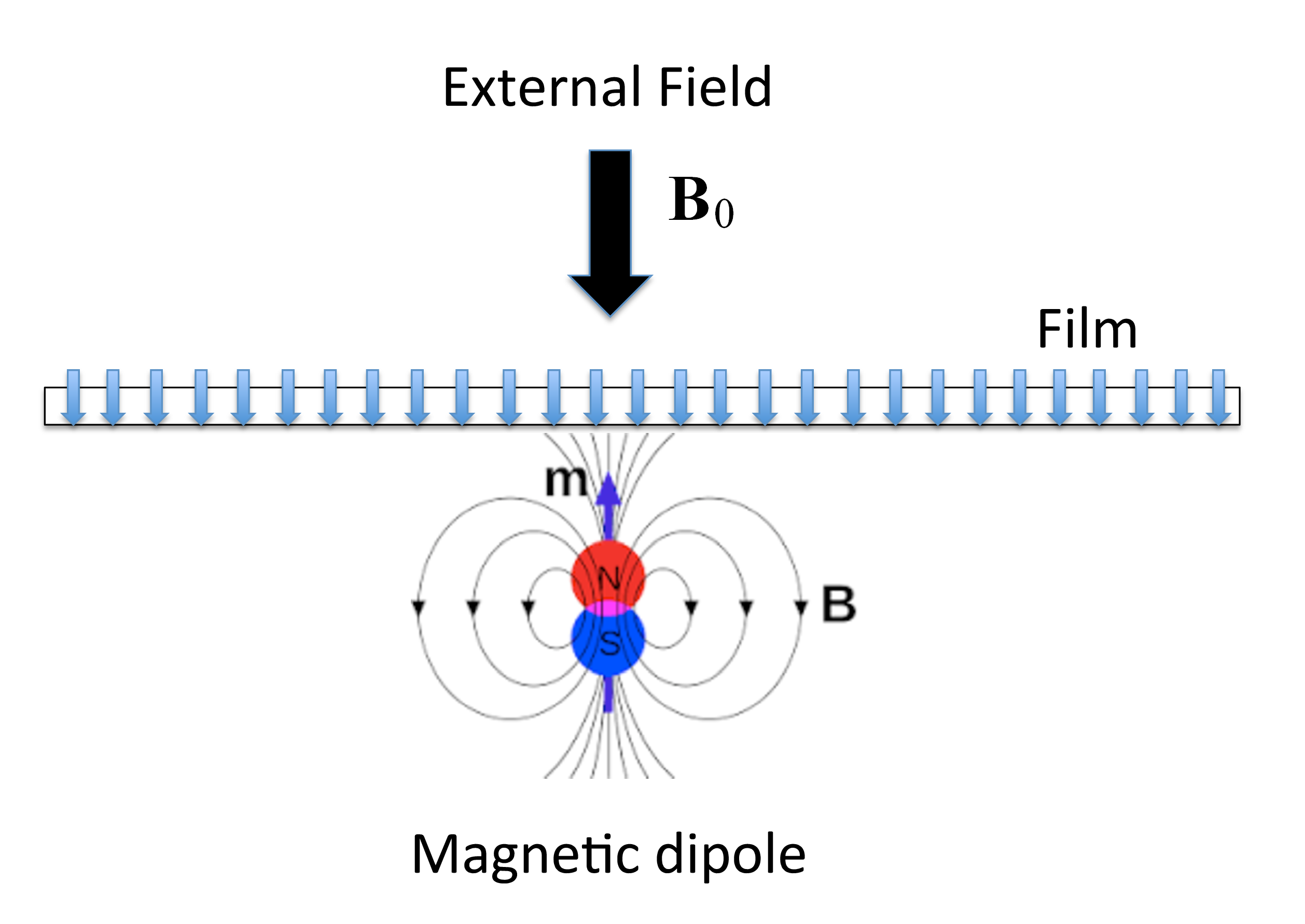} \caption{Geometry of the problem studied in the paper: A magnetic dipole with
the magnetic moment ${\bf m}$ approaches a film where the exchange-coupled
spins are aligned perpendicular to the film by the external field
$\mathbf{B}_{0}$. At some critical distance to the film, the dipole
nucleates a skyrmion by inducing local reversal of the spin field.
As will be seen in the computation, the initial bifurkation occurs
with a conservation of the topological charge, $Q=0$, by nucleating
a skyrmion-antiskyrmion pair. By moving closer to the film the dipole
forces the antiskyrmion to collapse, leaving behind a non-zero topological
charge $Q=1$ of the remaining skyrmion.}
\label{Fig_geometry} 
\end{figure}

Recently, it has been experimentally demonstrated and confirmed through
micromagnetic computations that stripe domains in a film can be cut
into skyrmions by the magnetic field of the tip of a scanning magnetic
force microscope (MFM) \cite{Senfu-APL2018}. In this paper we are
asking whether the field of a nanoscale magnetic dipole can nucleate
a skyrmion in a controllable manner in a uniformly magnetized film,
see Fig. \ref{Fig_geometry}. We find that it is definitely possible
but, probably, not with the use of a typical MFM tip, which is too
small to provide enough Zeeman energy to nucleate a skyrmion in a
typical ferromagnetic film. Instead one should use greater-size magnetic
nanoparticles of the kind used in nanocantilevers for mechanical magnetometry
\cite{deLoubens}.

For a rough estimate, consider a magnetic dipole of the average size
$R$ at a distance $h\lesssim R$ from a 2D film when it will generate
the highest field in the film. Let $2E_{Z}$ be the gain in the Zeeman
energy per spin of the film due to the local reversal of the spin-field
by the field of the dipole. That reversal would generally occur in
the area of linear size $R$, providing the total energy gain of order
$4\pi E_{Z}(R/a)^{2}$, where $a$ is the lattice constant. To nucleate
a skymion the gain in the Zeeman energy must overcome the ground state
exchange energy of the skyrmion, $4\pi J$, where $J$ is the exchange
energy per spin of the film. Equating the two energies one obtains
$R/a\sim\sqrt{J/E_{Z}}$. The ratio $J/E_{Z}$ would typically be in
the ballpark of $10^{4}-10^{6}$. Thus the required size of the dipole
is likely to be over $30$nm, that is, greater than the typical curvature
radius of a modern MFM tip.

The above estimate is confirmed by our simulations and analytical
results presented below. However, the manner in which skyrmions are
nucleated by the magnetic dipole turns out to be more complicated
than a simple reversal of the spin-field in a finite area of the film.
The complification is due to the fact that the topological charge
of the spin-field cannot be trivially changed from $Q = 0$ in the uniformly
magnetized film to $Q = 1$ in the presence of the skyrmion. Consequently,
as is seen in our numerical experiment, nucleation of the skyrmion goes
through a few non-trivial stages. In the first stage the magnetic
dipole, on approaching the film, nucleates a skyrmion-antiskyrmion
pair with zero topological charge. Depending on parameters the pair
can be either separated in space or the antiskyrmion can be centered
inside the skyrmion in a donut-like structure. In the second stage,
as the dipole continues to approach the film, the antiskyrmion collapses
(or is pushed out of the donut and then collapses), leaving behind
the non-zero topological charge of the skyrmion.

It is important to emphasize that the above dynamics of the nucleation
of a skyrmion by the magnetic dipole with the change of topology would
not exist within continuous 2D spin-field exchange model that conserves
topological charge. For that reason, instead of using micromagnetic
theory, our simulations have been done by minimizing the energy of
interacting spins in a large square lattice. In this case, which resembles
experiments with real materials, the presence of the finite lattice
spacing, $a$, breaks the scale invariance of the 2D exchange interaction
that is responsible for the conservation of the topological charge
\cite{CCG-PRB2012}. Still the topological charge remains conserved
with good accuracy for spin structures that are large compared to
the lattice spacing, which corresponds to the continuous limit. By
looking how the structures evolve down to the lattice scale we have
been able to observe the abrupt change of the topological charge from
zero to one when the collapsing antiskyrmion reaches the atomic size.

This paper is organized as follows. The model and numerical method
are explained in Section \ref{Sec_method}. Numerical results on the
creation of skyrmions in non-chiral films are presented in Section
\ref{Sec_nonchiral}. In Section \ref{Sec_boundary} we consider creation
of skyrmions by the magnetic dipole at the boundary of the film. Nucleation
of skyrmions by a magnetic dipole in a chiral system with the DMI
is discussed in Section \ref{Sec_chiral}. Analytical model that agrees
with numerical results is presented in Section \ref{Sec_analytical}.
Our results, numbers, and suggestions for experiments are discussed
in Section \ref{Sec_Discussion}.

\section{The model and numerical method}

\label{Sec_method}

We consider the Hamiltonian 
\begin{equation}
{\cal H}={\cal H}_{s}-\sum_{i}{\bf s}_{i}\cdot({\bf B}_{0}+{\bf B}_{di}),
\end{equation}
where the first term represents spin-spin interactions in a 2D lattice
and the second term represents Zeeman interaction of the spins with
the magnetic field. The latter consists of a constant external transverse
field, ${\bf b}_{0}$, and the field of the magnetic dipole, ${\bf b}_{d}$,
with ${\bf B}_{0}=g\mu_{B}S{\bf b}_{0}$ and ${\bf B}_{d}=g\mu_{B}S{\bf b}_{d}$
being the corresponding Zeeman energies per spin $S$ of the unit cell of the film and $g$
being the gyromagnetic factor associated with $S$.

We approximate the magnetic dipole by a point magnetic moment, ${\bf m}=m\mathbf{e}_{z}$,
positioned at the distance $h$ below the film and directed opposite
to the magnetization of the film, see Fig.\ \ref{Fig_geometry}.
The field of the dipole is given by 
\begin{equation}
{\bf b}_{d}({\bf r})=\frac{\mu_{0}}{4\pi}\left[\frac{3{\bf r}({\bf m}\cdot{\bf r})}{r^{5}}-\frac{{\bf m}}{r^{3}}\right],
\end{equation}
where ${\bf r}$ is the radius-vector originating at the dipole. Writing
for the points of the film ${\bf r}=(x,y,h)$, with $r=\sqrt{\rho^{2}+h^{2}}$
and $\bm{\rho}=(x,y)$, one has for the components of the dipole field
in the film 
\begin{eqnarray}
b_{dx} & = & \frac{\mu_{0}m}{4\pi}\frac{3hx}{(\rho^{2}+h^{2})^{5/2}}\\
b_{dy} & = & \frac{\mu_{0}m}{4\pi}\frac{3hy}{(\rho^{2}+h^{2})^{5/2}}\\
b_{dz} & = & \frac{\mu_{0}m}{4\pi}\frac{2h^{2}-\rho^{2}}{(\rho^{2}+h^{2})^{5/2}}
\end{eqnarray}
We used discretized version of these expressions to obtain ${\bf B}_{di}$
acting on the $i$-th spin in the film.

The field of the dipole at the closest point in the film, ${\bf r}=(0,0,h)$,
that equals 
\begin{equation}
b_{h}=\frac{\mu_{0}m}{2\pi h^{3}},\quad B_{h}=\frac{gS\mu_{0}\mu_{B}m}{2\pi h^{3}}\label{Bh}
\end{equation}
has been used to form a dimensionless parameter $B_{h}/(JS^{2}).$
Its value at a fixed $h$ depends on the magnetic moment, $m$, of
the dipole. At a given $h$ and $B_{0}$ we find the critical values
of $B_{h}/(JS^{2})$ that correspond to each stage of the nucleation
of the skyrmion.

Our numerical method, that is described in detail in Ref. \onlinecite{garchupro13},
consists of the minimization of the total energy of interacting spins
in a square latice of size up to $500\times500$. It involves successive
rotations of spins at lattice sites $i$ in the direction of the effective
field $\mathbf{H}_{\mathrm{eff},i}=-\delta{\cal H}/\delta{\bf S}_{i}$
(with ${\cal H}$ being the Hamiltonian of the system) with the probability
$\alpha$ and \textit{overrelaxation} (i.e., flipping spins around
$\mathbf{H}_{\mathrm{eff},i}$) with the probability $1-\alpha$.
The first operation reduces the energy of the system while the second
serves to better explore the phase space of the system via conservative
pseudo-dynamics, with $\alpha$ playing the role of the relaxation
constant. The fastest energy minimization towards the deepest minimum
is achieved for $\alpha\ll1$. We use $\alpha=0.01$.

Together with computing the spin configuration that minimizes the
energy, we also compute topological charge by using discretized form
of the expression 
\begin{equation}
Q=\int\frac{d^{2}\rho}{8\pi}\epsilon_{\alpha\beta}s_{a}\epsilon_{abc}\frac{\partial s_{b}}{\partial\rho_{\alpha}}\frac{\partial s_{c}}{\partial\rho_{\beta}}=\int\frac{dxdy}{4\pi}\:{\bf s}\cdot\frac{\partial{\bf s}}{\partial x}\times\frac{\partial{\bf s}}{\partial y}.\label{Q}
\end{equation}
Skyrmions have $Q=1$ while antiskyrmions have $Q=-1$. The skyrmion
size $\lambda$ has been extracted from the numerical data as \cite{CCG-PRB2012}
\begin{equation}
\lambda_{m}^{2}=\frac{m-1}{2^{m}\pi}a^{2}\sum_{i}\left(s_{iz}+1\right)^{m},
\end{equation}
with $s_{iz}=-1$ in the background and $s_{iz}=1$ at the center
of the skyrmion. For Belavin-Polyakov skyrmions \cite{BP} one has
$\lambda_{m}=\lambda$ for any $m$. We used $\lambda_{\mathrm{eff}}=\lambda_{4}$
to represent skyrmion size computed numerically.

\section{Nucleation of non-chiral skyrmions by a magnetic dipole}

\label{Sec_nonchiral}

\begin{figure}[ht]
\centering{}\includegraphics[width=8cm]{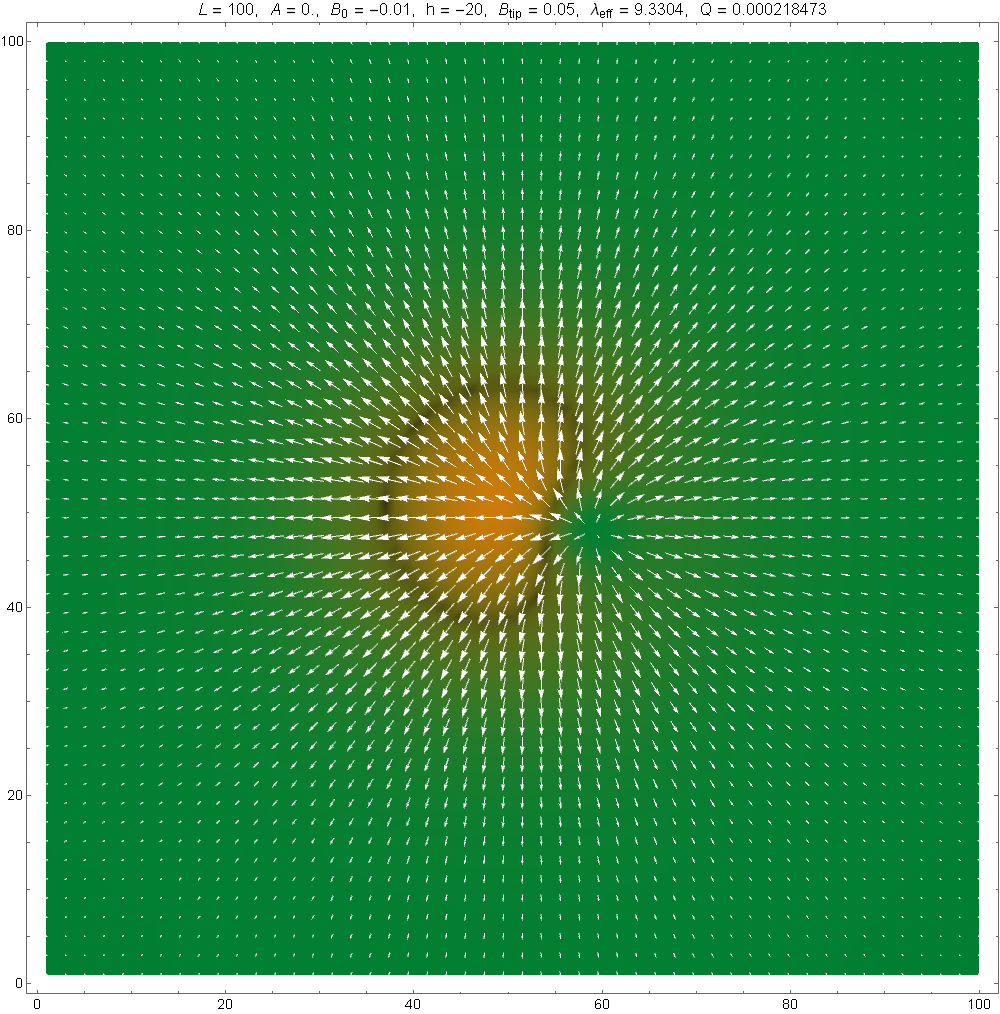} \includegraphics[width=8cm]{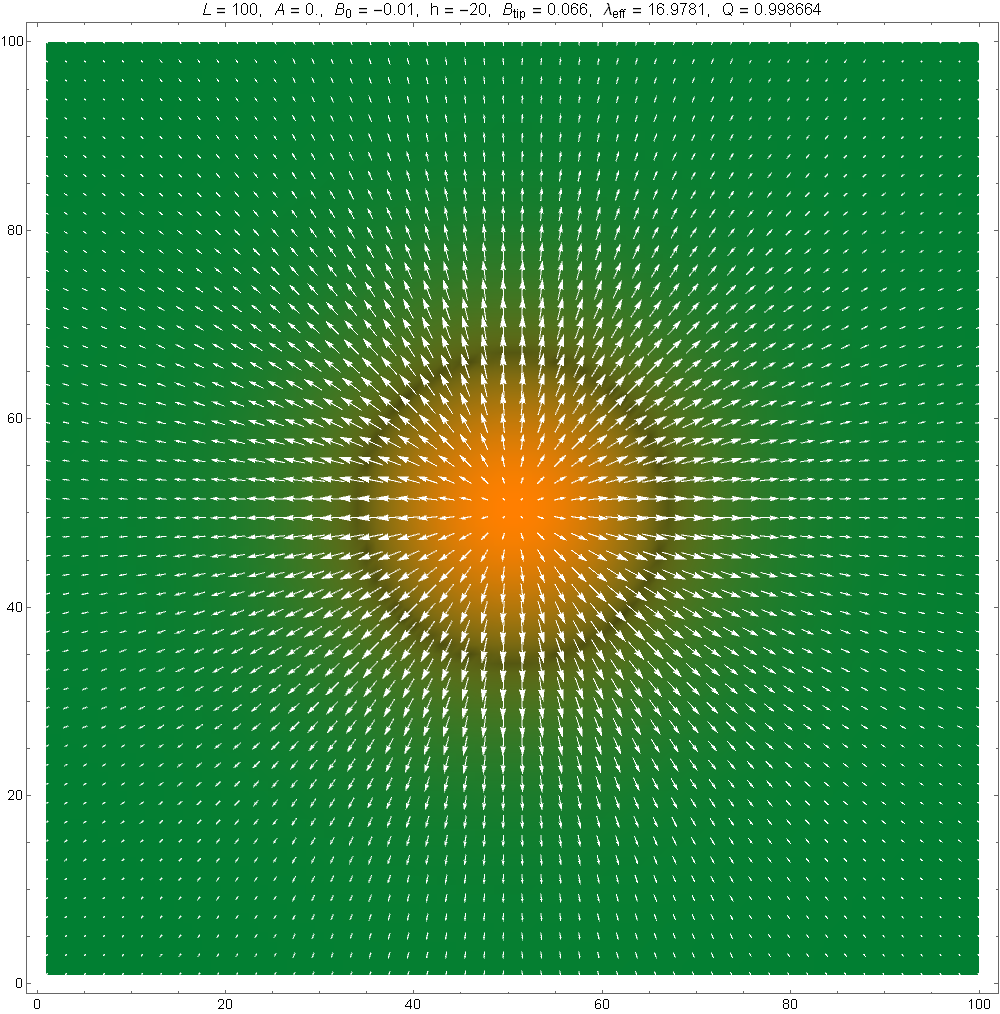}
\caption{Stages of the nucleation of a skyrmion by the magnetic dipole. The
downward magnetization is shown by green while the upward magnetization
is shown by orange. White arrows represent the in-plane spin components.
Upper panel: At $B_{h}=B_{h,1}=B_{h,2}$, the spin configuration with
all spins down becomes unstable and an asymmetric skyrmion-antiskyrmion
pair is formed. Lower panel: As $B_{h}$ increases, the skyrmion grows,
while the antiskyrmion shrinks and eventually collapses at $B_{h}=B_{h,Q}$.
Here the topological charge $Q$ of the spin configuration abruptly
changes from $0$ to $1$. }
\label{Fig_scenario_1} 
\end{figure}

In this section we consider the simplest case of the spin Hamiltonian,
\begin{equation}
{\cal H}_{s}=-\frac{S^{2}}{2}\sum_{ij}J_{ij}{\bf s}_{i}\cdot{\bf s}_{j},
\end{equation}
where $J_{ij}$ is the nearest-neighbor exchange interaction with
the coupling constant $J$. In numerical work we use $s=1$ and incorporate
the spin of the lattice site $S$ into the exchange constant $JS^{2}\rightarrow J$.

In the computations, a downward stabilizing field $B_{0}\ll J$ was
applied, so that $s_{z}\cong-1$ far from the magnetic dipole, the
distance $h$ was fixed and $B_{h}$ was increased in small steps
starting from zero, at each step minimizing the energy of the system.
The maximum value of $s_{z}$ was monitored. The value of $B_{h}$
at which $s_{z,\max}$ became positive was recorded as $B_{h,1}$.
Also the value $s_{z}=s_{z,\mathrm{center}}$ at $\rho=0$ (just above
the magnetic dipole) was monitored. The value of $B_{h}$ at which
$s_{z,\mathrm{center}}$ became positive was recorded as $B_{h,2}$.
The value of $B_{h}$ at which $Q=0$ changed to $Q=1$ (creation
of a skyrmion) was recorded as $B_{h,Q}$. 

The computations could also be done by approaching the magnetic dipole
to the film, i.e., keeping $m=\mathrm{const}$ and decreasing the
distance $h$ that also leads to the increasing of $B_{h}$. This
would better reflect real experiments but the method described above is more convenient
numerically as the region of the film influenced by the magnetic dipole
is constant. The phase diagram of critical parameters at which the
skyrmion is created can be recomputed in any desirable form for a concrete experiment.

In the first scenario illustrated in Fig.\ \ref{Fig_scenario_1}, at a certain value of $B_{h}$
the instability of the spin configuration in which all spins look down is observed: The magnetization of the film becomes
inverted near $\rho=0$ with a formation of the asymmetric skyrmion-antiskyrmion
pair. In this scenario $B_{h,1}=B_{h,2}$. Further increase of $B_{h}$
leads to the collapse of the antiskyrmion and the abrupt change of the topological
charge from $0$ to $1$ at $B_{h}=B_{h,Q}$.

\begin{figure}[ht]
\centering{}\includegraphics[width=8cm]{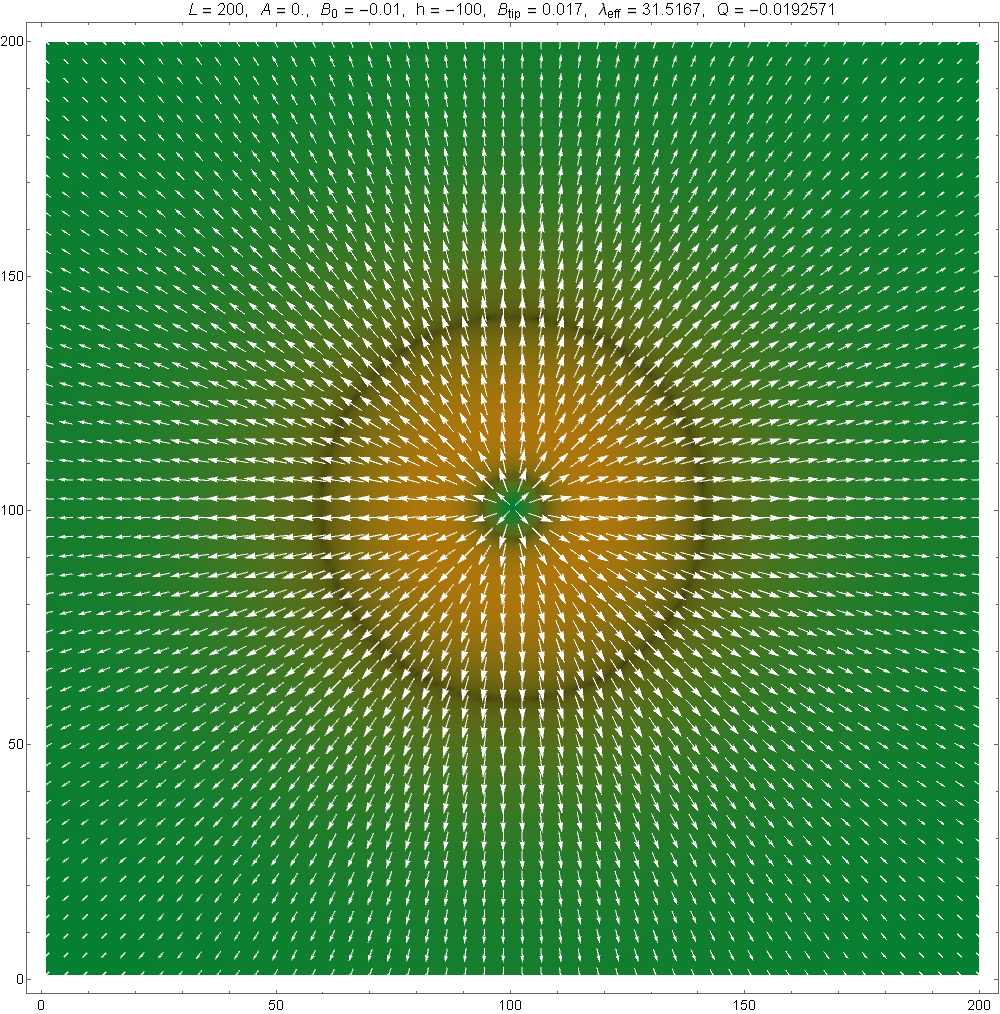} \caption{Skyrmion-antiskyrmion donut with $Q=0$ formed by the magnetic dipole
on approaching the film.}
\label{Donut} 
\end{figure}

In the second scenario, a cylindrically symmetric donut with $Q=0$,
containing an antiskyrmion inside a skyrmion, see Fig.\ \ref{Donut},
is formed via continuous rotation of the spins at $\rho\sim h$ under
the combined influence of the vertical and in-plane components of
the magnetic field created by the magnetic dipole (see Fig. \ref{Fig_geometry}).
For the donut one still has $s_{z,\mathrm{center}}\cong-1$, thus $B_{h,1}<B_{h}<B_{h,2}$.
Upon further increasing $B_{h}$, the outer radius of the donut increases
while its inner radius representing the size of the antiskyrmion decreases. At some
$B_{h}=B_{h2}=B_{h,Q}$, the antiskyrmion collapses leaving only a skyrmion
with $Q=1$ in the film. 

\begin{figure}[ht]
\centering{}\includegraphics[width=9cm]{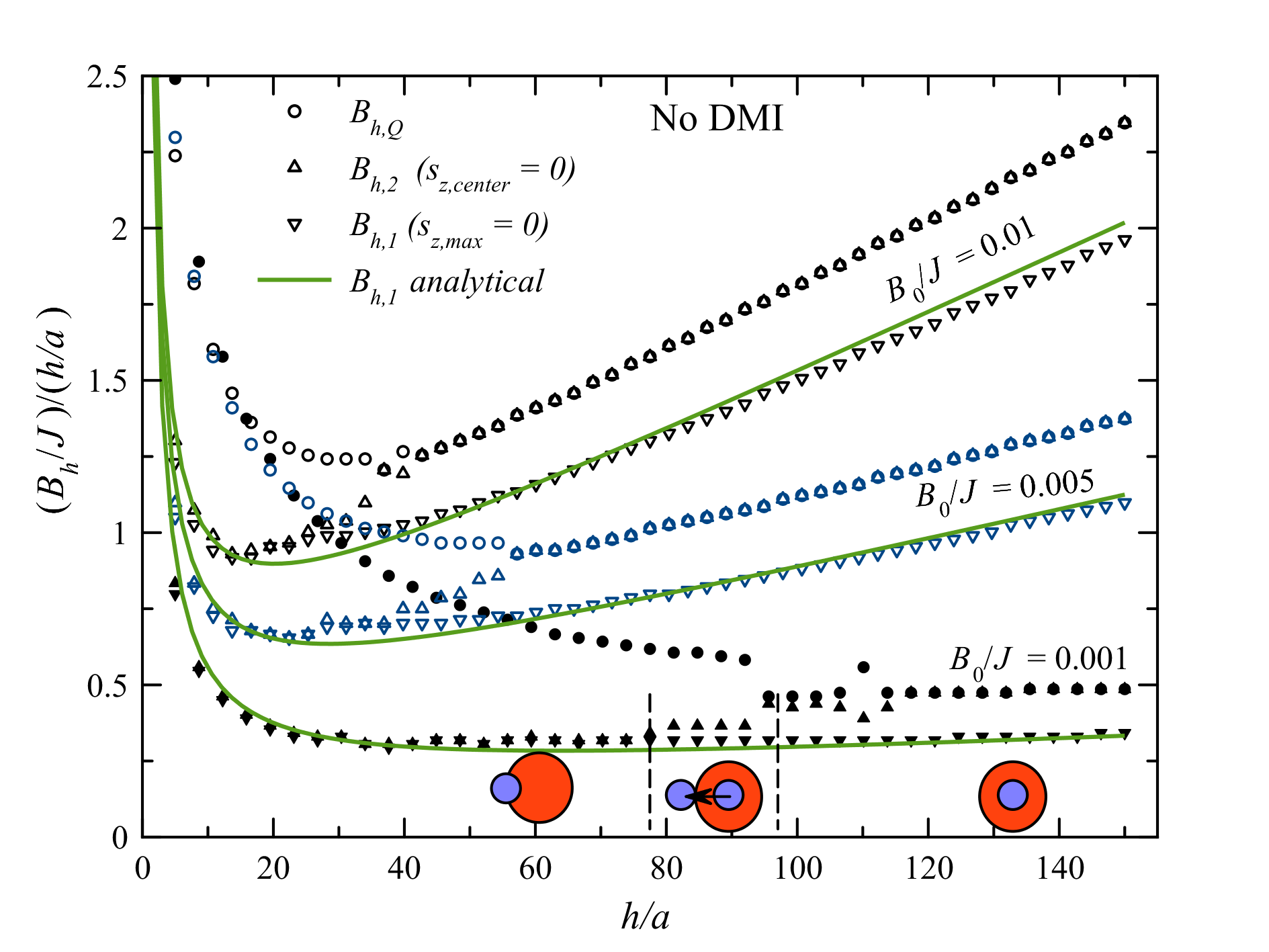} \caption{Dependence of the critical fields $B_{h,1},B_{h,2},B_{h,Q}$, see
text, on the distance, $h$, of the dipole to the film for different
values of $B_{0}$. Solid lines show theoretical curves computed in
Section \ref{Sec_analytical}.}
\label{Fig_NoDMI_PHD} 
\end{figure}

The skyrmion-nucleation phase diagram containing critical branches
of $B_{h}(h)$ (multiplied by $h/a$ for better presentation) is shown
in Fig.\ \ref{Fig_NoDMI_PHD}. The first scenario is realized for
smaller $h$ while the second scenario is realized for larger $h$.
There is a relatively narrow region of $h$ in which a combined scenario
with $B_{h,1}<B_{h,2}<B_{h,Q}$ is realized. Here, first a donut is
created and then it loses its symmetry via expulsion of the antiskyrmion
to the periphery of the skyrmion, where it collapses upon further increase of
$B_{h}$. All scenarios are shown schematically near the bottom of
the figure.

\begin{figure}
\centering{}\includegraphics[width=9cm]{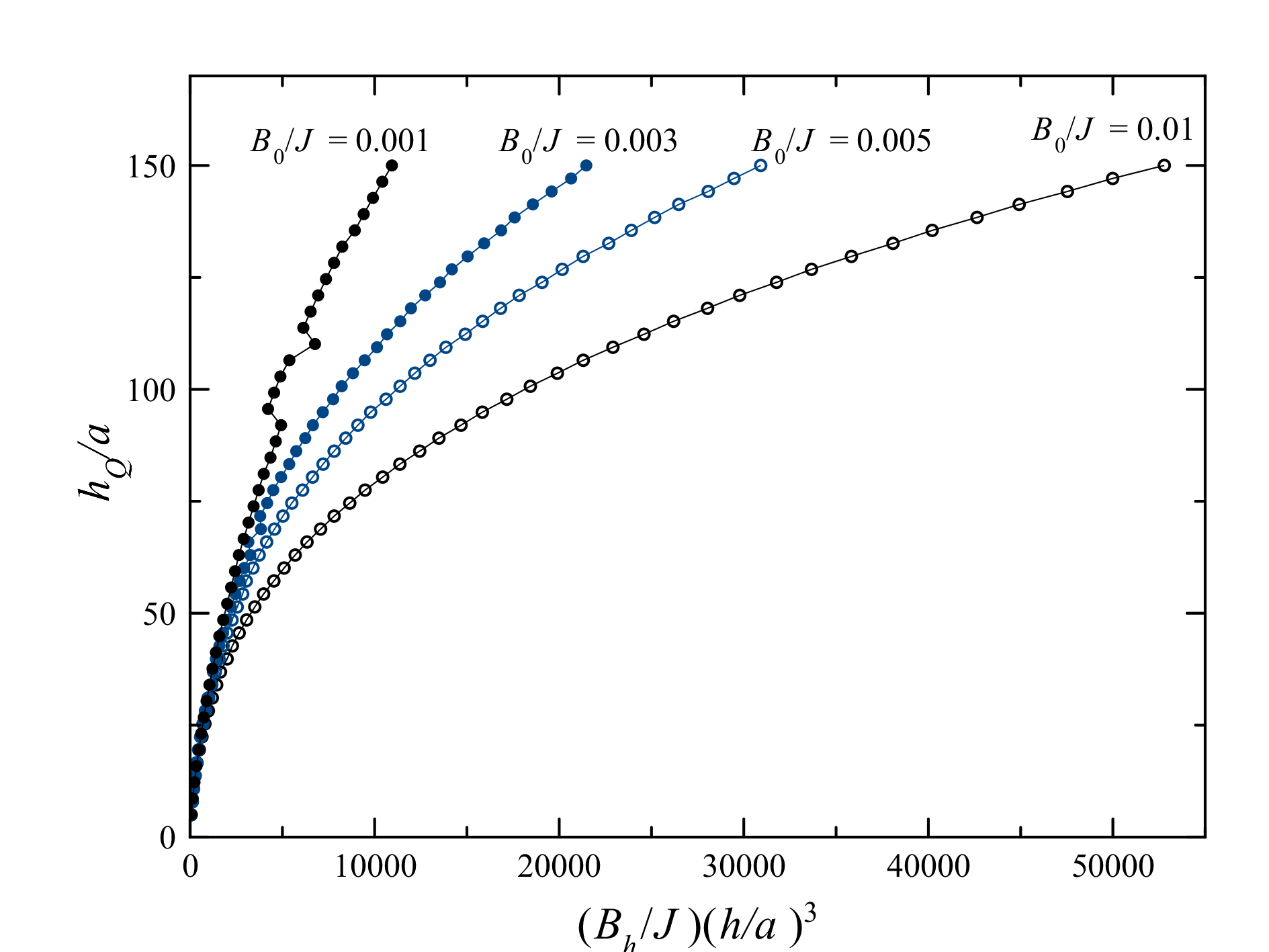}\caption{Critical distance of the magnetic dipole from the film, $h_{Q}$,
at which the skyrmion is created, vs $\left[B_{h}/\left(JS^{2}\right)\right](h/a)^{3}$
that is proportional to the magnetic moment, $m$, of the dipole.\label{Fig_hQ_vs_m}}
\end{figure}

The same data are represented in Fig.\ \ref{Fig_hQ_vs_m} in the form of the dependence of $h_{Q}$ (the distance at which the skyrmion is created) on
$\left[B_{h}/\left(JS^{2}\right)\right](h/a)^{3}=gS\mu_{0}\mu_{B}m/\left(2\pi a^{3}JS^{2}\right)\propto m$,
see Eq. (\ref{Bh}). This figure corresponds to the experimental
situation in which a magnetic dipole of a fixed strenth $m$ is approaching
the film.


\section{Nucleation of skyrmions at the edge of the film}

\label{Sec_boundary}

\begin{figure}[ht]
\centering{}\includegraphics[width=9cm]{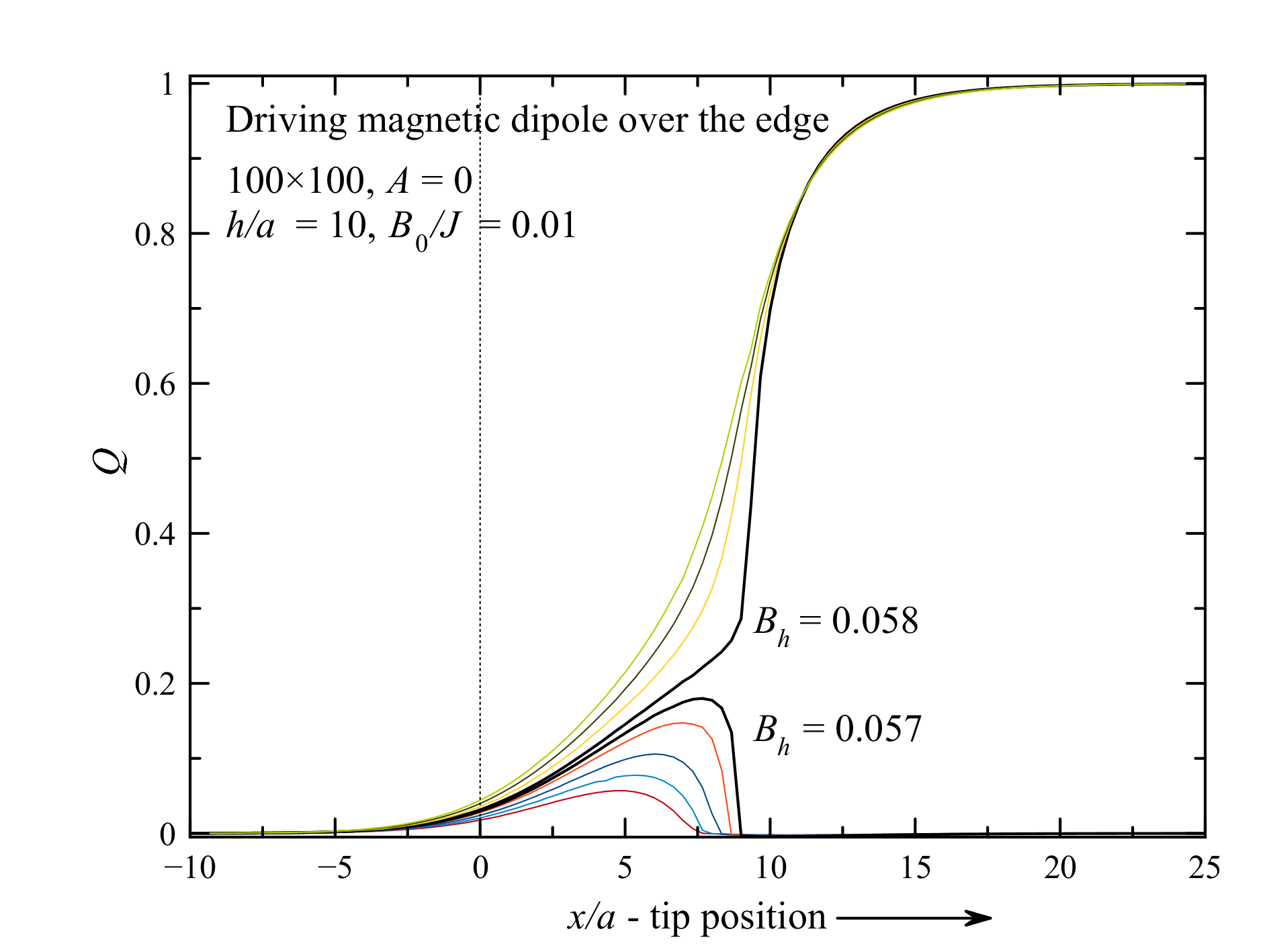} \caption{Evolution of the topological charge in a non-chiral film in the process
of skyrmion creation by the magnetic dipole moving along the $x$-axis parallel to the
film at $h=10a$  and crossing its boundary at $x=0$. When $B_{h}$ is above a certain threshold indicated in the figure, $Q$ changes from 0 to 1 as the dipole moves through a distance of a few $h$.}
\label{Fig_tip_driving} 
\end{figure}

\begin{figure}
\begin{centering}
\includegraphics[width=9cm]{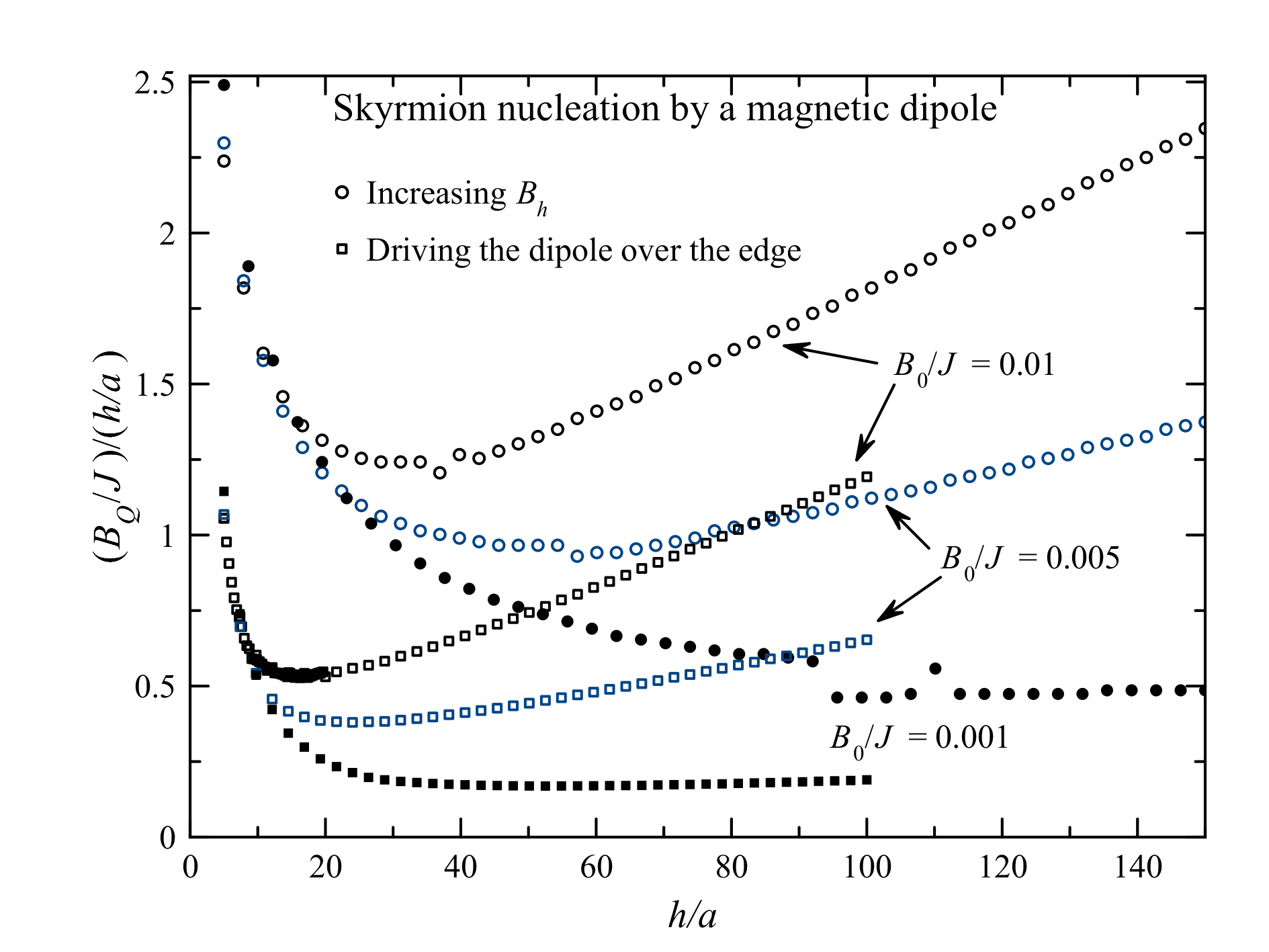}
\par\end{centering}
\caption{Phase diagram for the skyrmion nucleation by the magnetic dipole: Changing
$B_{h}$ (the data from Fig. \ref{Fig_NoDMI_PHD}) vs driving the
dipole over the film's edge. The latter can be achieved with a smaller magnetic moment of the dipole.}

\label{Fig_PHD}

\end{figure}

Here we study the nucleation of the skyrmion by the magnetic dipole
moving parallel to the film and crossing its boundary from outside,
starting at $x<0$ at a distance satisfying $|x|\gg h$. This method is more efficient
than the method considered above as it requires a smaller dipole field
$B_{h}$ for the skyrmion nucleation. As the magnetic dipole is approaching
the edge of the film and crossing it  at $x=0$, the topological charge
$Q$ is gradually increasing from zero (see Fig. \ref{Fig_tip_driving}).
Due to the boundary, close to it $Q$ is not quantized and can take any value
$0\leq Q\leq1$. At $x\sim h$, there is a bifurcation: If $B_{h}$
is too weak, the skyrmion is not created and $Q$ quickly returns
to zero. When $B_{h}$ exceeds a certain threshold, the skyrmion
is created and $Q$ approaches 1 as the magnetic dipole continues to move
above the film. The bifurcation
value of $B_{h}$ is recorded as $B_{h,Q}$. 

The resulting values of $B_{h,Q}$ (multiplied by $h/a$) are represented
in Fig. \ref{Fig_PHD} together with the data obtained in the previous
section by increasing $B_{h}$. One can see that in terms of the required magnetic moment the method based upon driving
the dipole over the edge parallel to the film is more efficient than the method based upon moving the dipole in the direction normal to the film far from edges. 

\section{Nucleation of chiral skyrmions by a magnetic dipole}

\label{Sec_chiral}

\begin{figure}
\centering{}\includegraphics[width=75mm]{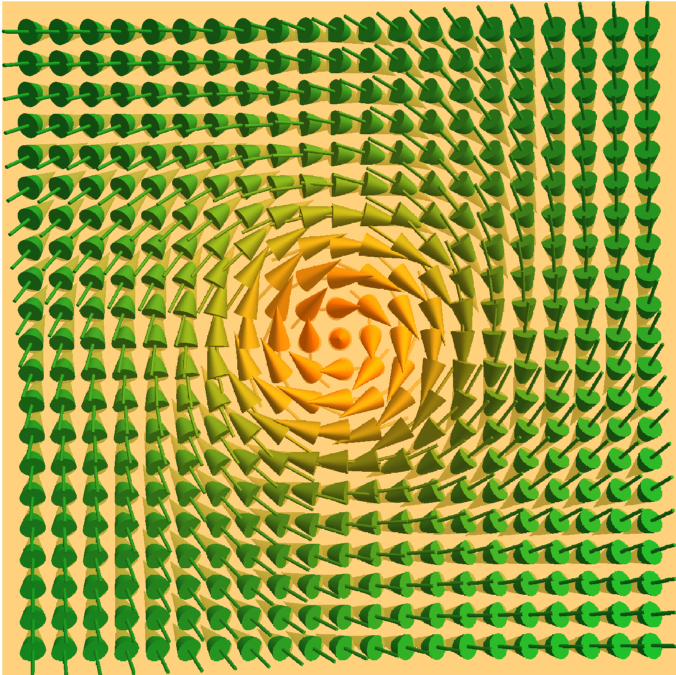} \includegraphics[width=75mm]{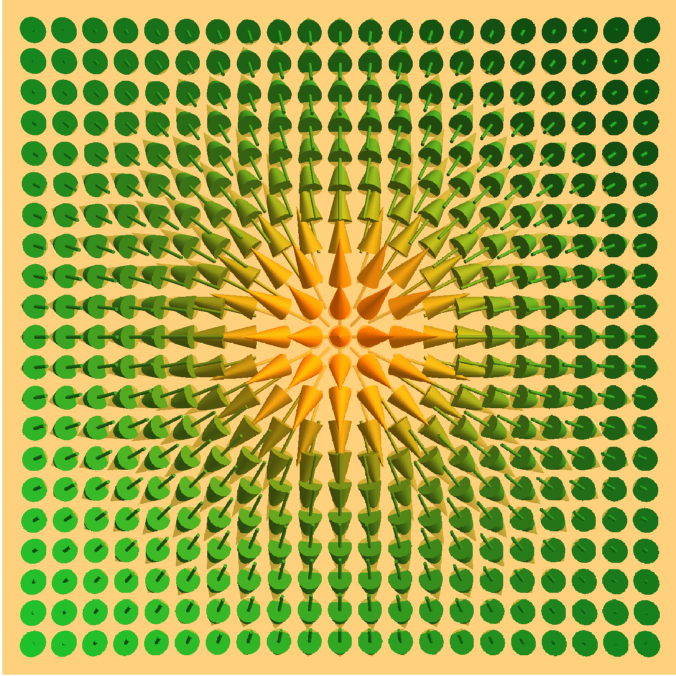}
\caption{Spin field of the Belavin-Polyakov skyrmions, $Q=1$. Upper panel:
Bloch-type, counterclockwise for $A<0$; Lower panel: N\'eel-type, outward
for $A>0$. }
\label{BP-skyrmions} 
\end{figure}

In this section we consider a film with the Dzyaloshinskii-Moriya
interaction (DMI) and add 
\begin{equation}
{\cal H}_{\mathrm{DMI}}=A\sum_{i}[({\bf S}_{i}\times{\bf S}_{i+\hat{x}})\cdot\mathbf{e}_{x}+({\bf S}_{i}\times{\bf S}_{i+\hat{y}})\cdot\mathbf{e}_{y}]
\end{equation}
to the exchange interaction. This Hamiltonian describes Bloch-type
DMI of strength $A$ in a non-centrosymmetric crystal \cite{Leonov-NJP2016}.
For the N\'eel-type DMI it should be replaced with $A\sum_{i}[({\bf S}_{i}\times{\bf S}_{i+\hat{x}})\cdot\mathbf{e}_{y}-({\bf S}_{i}\times{\bf S}_{i+\hat{y}})\cdot\mathbf{e}_{x}]$.
The spin-fields in the N\'eel-type ($\gamma=0$) and Bloch-type ($\gamma=\pi/2$)
skyrmions are shown in Fig.\ \ref{BP-skyrmions}.

In the case of $A\ll J$, the DMI only insignificantly changes the skyrmion
nucleation condition. For stronger DMI, there is a difference for
different types of the DMI, Bloch or N\'eel, and for different signs of $A$ in
the N\'eel case. In the geometry shown in Fig. \ref{Fig_geometry},
the magnetic dipole creates the N\'eel-type skyrmion with an outward looking spin-field. Consequently, the N\'eel-type DMI with $A>0$ helps the skyrmion nucleation,
thus the corresponding values of $B_{h}$ are lower than in the pure-exchange
model. On the contrary, for $A<0$, the DMI works against the magnetic dipole and a greater $B_{h}$ is required to nucleate a skyrmion. For the Bloch-type DMI, the initial
instability happens early, so that $B_{h,1}$ is lower than in the
pure-exchange model. However, it is difficult to finish the process
and create a skyrmion because $B_{h,Q}$ is significantly higher than
in the pure-exchange model. Thus, a strong Bloch-type DMI is undesirable
for the skyrmion creation by the magnetic dipole.

Notice that in the absence of the stabilizing field $B_{0}$ the DMI favors a laminar domain
structure even in the absence of the DDI. Thus, the stronger DMI, the stronger $B_{0}$ is required to create a uniformly magnetized state. This, in turn, requires a stronger magnetic dipole to nucleate a skyrmion, making strong DMI of any type unfavorable for this purpose. A special case is when $B_{0}$ is chosen such that the uniform state
is on the verge of stability. However, in a sample of finite dimensions the loss of stability of the uniformly magnetized film  on decreasing $B_0$ always occurs at the edges of the film, while in the middle the
uniform state remains rather stable. Driving the magnetic dipole parallel to the film and crossing its
boundary, that worked well for the pure-exchange model, may be also problematic
for a strong DMI. When the uniform state was on the verge of breaking into domains,
the moving magnetic dipole in our simulation was creating a trailing finger domain
instead of a skyrmion.

\begin{figure}[ht]
\centering{}\includegraphics[width=9cm]{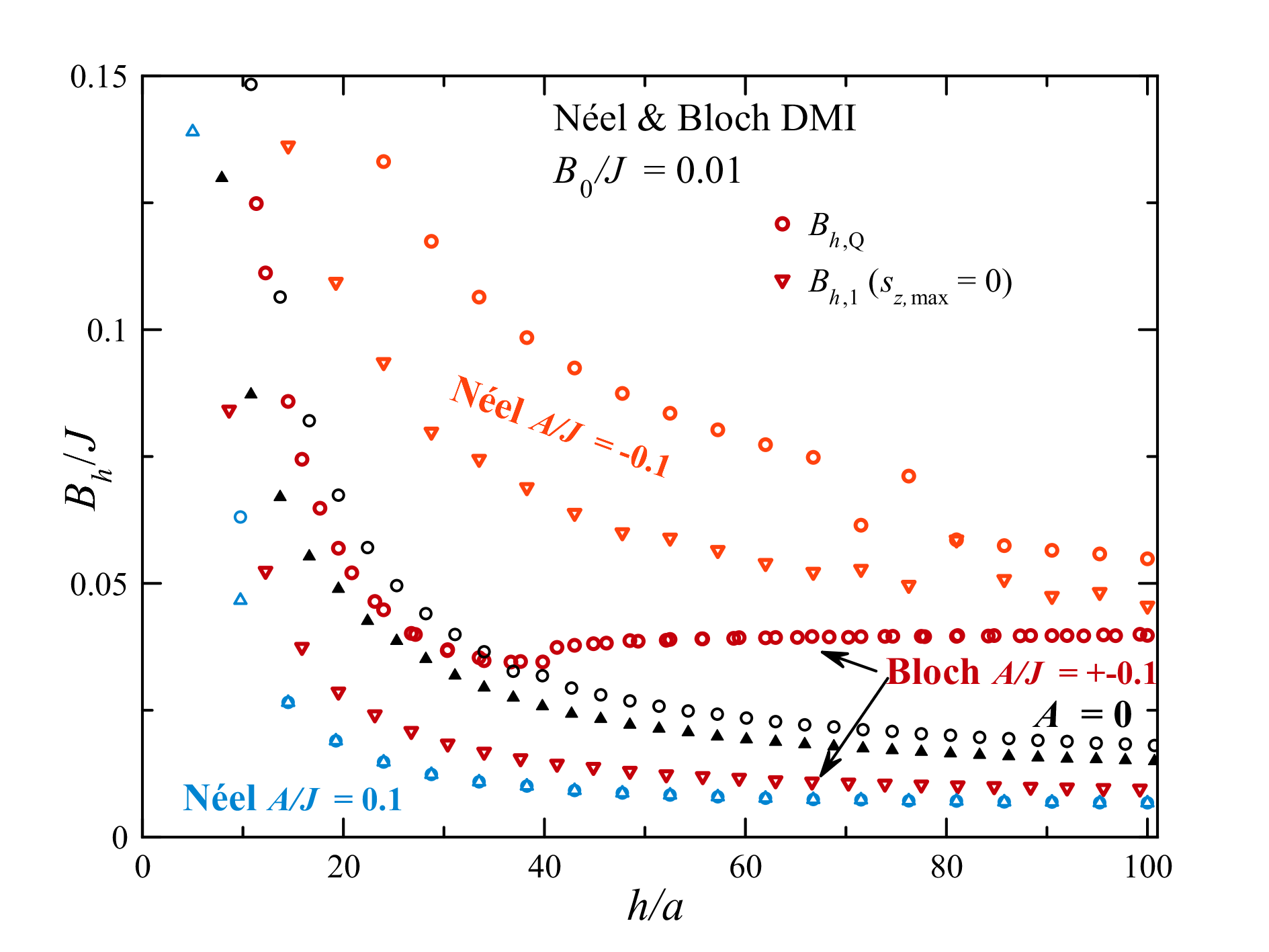} \caption{Dependence of the critical fields $B_{h,1}$ and $B_{h,Q}$ on the
distance of the dipole to the film for N\'eel and Bloch DMI at $B_{0}/J=0.01$
and $A/J=\pm0.1$.}
\label{Fig_chiral} 
\end{figure}

\section{Analytical model}

\label{Sec_analytical} In this section we develop analytical model
of the instability of the uniform state in the presence of the uniform
stabilizing field $B_{0}$ and the opposite field of the magnetic
dipole that explains quantitatively our findings for a non-chiral
film. This instability is due to the normal component of the dipole's
field, so we discard the in-plane components. Using a continuous spin-field
model obtained by replacing $\sum_{i}\Rightarrow\int d^{2}\rho/a^{2}$
and writing 
\begin{equation}
s_{z}=-\sqrt{1-s_{x}^{2}-s_{y}^{2}}\approx-1+\frac{1}{2}(s_{x}^{2}+s_{y}^{2}),
\end{equation}
one obtains the Zeeman energy due to the dipole as 
\begin{equation}
E_{d}=-\frac{B_{h}h^{3}}{4a^{2}}\int d^{2}\rho\frac{2h^{2}-\rho^{2}}{(\rho^{2}+h^{2})^{5/2}}(s_{x}^{2}+s_{y}^{2}),
\end{equation}
while Zeeman energy due to the external field is 
\begin{equation}
E_{0}=\frac{B_{0}}{2a^{2}}\int d^{2}\rho(s_{x}^{2}+s_{y}^{2}).
\end{equation}
The continuous counterpart of the exchange energy due to the development
of the tranverse components of the spin field is 
\begin{eqnarray}
E_{ex} & = & JS^{2}\int dxdy\left[\left(\frac{\partial s_{x}}{\partial x}\right)^{2}+\left(\frac{\partial s_{x}}{\partial y}\right)^{2}\right.\nonumber \\
 & + & \left.\left(\frac{\partial s_{y}}{\partial x}\right)^{2}+\left(\frac{\partial s_{y}}{\partial y}\right)^{2}\right]
\end{eqnarray}

One kind of instability observed in numerical experiment consists of tilting the spins in the vicinity of the dipole,
all in one direction. Without limiting generality, one can consider
the $x$-axis to be the direction of the tilt, i.e., $s_{y}=0$, $s_{x}=f(\rho)$,
where $f(\rho)$ is a trial function that we choose in the form 
\begin{equation}
f(\rho)=\frac{Ch^{2\alpha}}{(h^{2}+\rho^{2})^{\alpha}}\label{trialf}
\end{equation}
with $\alpha$ being an unknown exponent to be determined. This results
in the following expressions for the above energies: 
\begin{eqnarray}
E_{d} & = & -\frac{4\pi B_{h}C^{2}h^{2}}{a^{2}}\frac{\alpha}{(4\alpha+1)(4\alpha+3)}\\
E_{0} & = & \frac{\pi B_{0}C^{2}h^{2}}{2a^{2}}\frac{1}{2\alpha-1}\\
E_{ex} & = & 2\pi JS^{2}C^{2}\frac{\alpha}{2\alpha+1}.
\end{eqnarray}
Instability occurs when 
\begin{equation}
E_{d}+E_{0}+E_{ex}\leq0,
\end{equation}
with the instability threshold given by the equal sign. It
provides the critical value of the dipole's field $B_{h}(h,\alpha)$
that has to be minimized with respect to $\alpha$. The analysis is
facilitated by the reduced variables 
\begin{equation}
\tilde{B}_{h}\equiv\frac{B_{h}}{JS^{2}}\left(\frac{h}{a}\right)^{2},\qquad\tilde{B}_{0}\equiv\frac{B_{0}}{JS^{2}}\left(\frac{h}{a}\right)^{2}.
\end{equation}
For $\tilde{B}_{0}\ll1$, one has $\alpha$ close to 1/2 that simplifies
the analytics. In this region one obtains 
\begin{equation}
\tilde{B}_{h}\cong\frac{15}{4}+\sqrt{\frac{255}{8}\tilde{B}_{0}}.\label{small_B0}
\end{equation}
In the opposite limit $\tilde{B}_{0}\gg1$, one has $\alpha\gg1$
and the minimization simplifies again, leading to
\begin{equation}
\tilde{B}_{h}\cong\tilde{B}_{0}+2\sqrt{6\tilde{B}_{0}}.\label{large_B0}
\end{equation}
The two limiting formulas above can be combined into one formula
\begin{equation}
\tilde{B}_{h}\cong\frac{15}{4}+\sqrt{\frac{255}{8}\tilde{B}_{0}}.+\frac{\tilde{B}_{0}^{3/2}}{\sqrt{\tilde{B}_{0}}+\sqrt{255/8}-2\sqrt{6}}\label{Bhtilde_combined}
\end{equation}
that is practically indistinguishable from the result of the numerical
minimization of $B_{h}(h,\alpha)$. Eq. (\ref{Bhtilde_combined})
has been used to plot theoretical solid lines in Fig. \ref{Fig_NoDMI_PHD}. They are in a very good accord with the numerical result for $B_{h,1}$.
The region on the right in Fig. \ref{Fig_NoDMI_PHD} described by
Eq. (\ref{large_B0}) for $\tilde{B}_{0}\gg1$ is the most important
one because $B_{0}$ must be sufficiently large to prevent the magnetization
of the film from breaking into magnetic domains and because of the
limitation on the value of the magnetic moment of the dipole that
requires $h/a\gg1$. 

Critical fields corresponding to other stages of the nucleation process,
that occur in a strongly non-uniform magnetization phase, are more
difficult to obtain analytically, although by order of magnitude they
are in the same ballpark as the first critical field. The model with
the DMI turns out to be more challenging than the non-chiral model. Contributions from the DMI from the trial function of Eq. (\ref{trialf}) vanish, pointing to a more complex instability mode. 

\section{Discussion}

\label{Sec_Discussion} One necessary condition of nucleating a skyrmion
is that the field of the dipole exceeds the external field stabilizing
the uniform state, see Eq.\ (\ref{large_B0}). In the numerical and
analytical work we treated the magnetic dipole at a distance $h$
from the film as a point particle. It is clear, however, that by order
of magnitude all our results must be correct for a dipole of size
$R\sim h$. In fact $h\sim R$ would be best for providing the highest
field of the dipole in the film. Skyrmion nucleated by such a dipole
must be of a size $\lambda\sim h\sim R$.

To estimate the dimensions of skyrmions that can be nucleated in a
2D film by a magnetic dipole one has to equate $B_{h}$ determined
by Eqs.\ (\ref{small_B0}) or (\ref{large_B0}) to the field of the
dipole given by Eq.\ (\ref{Bh}). In both cases one obtains $h/a\sim\lambda/a\sim(JS^{2}/B_{h})^{1/2}$.
In accordance with the qualitative reasoning presented in the Introduction,
 $JS^{2}/B_{h}$ is the ratio of the exchange energy and Zeeman energy of the dipole per spin of the film, which is typically in the ballpark
of $10^{4}-10^{6}$. This gives $\lambda/a\sim10^{2}-10^{3}$. It
does not mean, however, that a skyrmion of that size will remain in
the film after the dipole is moved away. The skyrmion created by the dipole will either collapse or evolve towards a certain equilibrium size depending on whether skyrmions of a stable size exist due to all interactions present in the film. 

Note that in the numerical work we studied $B_{h}/(JS^{2})$
and $B_{0}/(JS^{2})$ greater than the ratios typically achieved in
real experiments unless one works with a low exchange system at low
temperatures. For the reason explained above the smaller values of
these ratios would generate larger skyrmions whose study would require
computation on spin lattices of impractically large size. This, however, in no away reduces the applicability of our numerical results to real experiments because the latter would follow the same instability patterns and the same scaling with parameters. Analytical
formulas provided in the paper, which agree well with numerical results,
provide guidance for experiments with real films and real dipoles.

In our treatment we neglected a number of interactions that could
be important for stabilizing skyrmions nucleated by a magnetic dipole
but which play lesser role in the nucleation process. Among them are
dipole-dipole interaction (DDI) and magnetic anisotropy (crystal field).
In the first approximation the omission of the DDI is justified by
the necessity to apply an external field that prevents the system
from breaking into magnetic domains. Such a field, by definition, must
be greater than dipolar fields in the film and so should be the field
of the magnetic dipole used. We also have assumed that the magnetic
anisotropy field is small compared to the external field. Generalization
that takes into account the omitted interactions is straightforward
but it would make the problem much messier because the nucleation threshold
would depend on a greater number of parameters. For simplicity we
talked about a single atomic layer of spins. The generalization to
$n$ atomic layers consists of replacing the exchange constant $J$
with $Jn$ as long as the condition $an<h$ is satisfied.

Besides the principle possibility of writing and manipulating skyrmions
with a magnetic dipole, our other interesting finding is the manner
in which skyrmions are nucleated by the magnetic dipole. Stages of this
nontrivial process are governed by topology that prohibits the
change of the topological charge of the spin-field that is a smooth
function of coordinates. The latter is dictated by the exchange interaction,
which is the dominant interaction in the system. To change the topology
one needs to reverse a single spin with respect to its neighbors,
which costs large exchange energy. This is observed in the numerical
experiment. It shows that the instability begins with the formation
of the skyrmion-antiskyrmion pair carrying zero topological charge.
On further approaching the film the dipole forces the antiskyrmion
to collapse, abruptly changing the topological charge from zero to
one due to the remaining skyrmion.

The method of writing skyrmions proposed in this paper should not
be difficult to test in real experiments if one chooses parameters right in accordance with our suggestions. Besides its potential for applications, it must be also interesting to observe the non-trivial stages of skyrmion nucleation by the magnetic dipole seen in numerical experiments. 

\section{Acknowledgements}

This work has been supported by the grant No. OSR-2016-CRG5-2977 from
King Abdullah University of Science and Technology.

\end{document}